\documentstyle[fleqn]{article}
\parindent=20pt  \def\r#1{$^{[#1]}$} 
\def\E{\mbox{e}^+\mbox{e}^-} 

\title{Nuclear Effect in Higher-Dimensional\\
Factorial Moment Analysis of the $^{16}$O-, $^{32}$S- and
$^{197}$\hskip-2truept Au-Em\\
Interaction Data at 200, 60 and 11 A GeV/c}
\author{EMU-01 Collaboration}
\date{}

\begin{document}
\maketitle

M.I.~Adamovich$^{15}$, M.M.~Aggarwal$^{4}$, Y.A.~Alexandrov$^{15}$,
R.~Amirikas$^{19}$, N.P.~Andreeva$^{1}$,
F.A.~Avetyan$^{23}$, S.K.~Badyal$^{9}$, A.M.~Bakich$^{19}$, E.S.~Basova$^{20}$,
K.B.~Bhalla$^{8}$
% ...
\\

\newpage

\newcounter{lab}
\begin{list}%
{$^{\;\, \arabic{lab}}$}{\usecounter{lab}
\setlength{\itemsep}{0.0cm} \setlength{\parsep}{0.0cm}}
\item High Energy Physics Institute, Almaty, Kazakhstan
\item Institute of High Energy Physics, Academia Sinica, Beijing, China
\item Institute for Gravitation and Space Research, Bucharest, Romania
\item Department of Physics, Panjab University, Chandigarh, India
\item Department of Physics, Hunan Education Institute, Changsha, Hunan, China
\item Laboratory of High Energies, Joint Institute for Nuclear Research (JINR), Dubna, Russia
\item Institute of Nuclear Physics, Gatchina, Russia
\item Department of Physics, University of Rajasthan, Jaipur, India
\item Department of Physics, University of Jammu, Jammu, India
\item Department of Nuclear Physics and Biophysics, Safarik University, Kosice, Slovakia
\item Institute of Experimental Physics, Slovak Academy of Sciences, Kosice, Slovakia
\item Department of Physics, Shanxi Normal University, Linfen, Shanxi, China
\item Department of Physics, University of Lund, Lund, Sweden
\item F.B. Physik, Philipps University, Marburg, Germany
\item P N Lebedev Physical Institute, Moscow, Russia
\item V G Khlopin Radium Institute, St. Petersburg, Russia
\item Department of Physics, University of Washington, Seattle, Washington, USA
\item Department of Physics, Kon-Kuk University, Seoul, Korea
\item School of Physics, University of Sydney, Sydney, Australia
\item Lab. of Relativistic Nuclear Physics, Institute of Nuclear Physics, Tashkent, Uzbekistan
\item Lab. of High Energies, Physical-Technical Institute, Tashkent, Uzbekistan
\item Institute of Particle Physics, Hua-Zhong Normal University, Wuhan, Hubei, China
\item Yerevan Physics Institute, Yerevan, Armenia
\end{list}

\maketitle
\bigskip

\begin{center}\begin{minipage}{120mm}\begin{abstract}
The anomalous behavior of 2-dimensional factorial moment in nucleus-nucleus
collisions is studied in some detail using both mini-bias and central
collision data of $^{16}$O-, $^{32}$S- and $^{197}$Au-Em interactions from
EMU01 experiment. The correct value for the effective Hurst exponent in the
analysis of higher-dimensional factorial moment is found to be greater than
unity, showing clearly the existence of superposition effect in nucleus-nucleus
collisions.
\end{abstract}\end{minipage}\end{center}

\newpage

\noindent
{\bf I \ \ Introduction}

\vskip0.2cm
\noindent
The scaling property of factorial moment (FM) defined as
$$F_q(M)={\frac {1}{M}}\sum\limits_{m=1}^{M}{{\langle n_m(n_m-1)\cdots
(n_m-q+1)\rangle }\over {{\langle n_m \rangle}^q}} \eqno(1)$$
has been studied widely in various kind of collisions, from hadron-hadron,
$\E$ to nucleus-nucleus at various energies\r{1}, and a rich pattern
of phenomena has been obtained. In Eq.(1) $M$ is the
partition number of a phase space region $\Delta$, $n_m$ is the multiplicity
in the $m$th subdivision cell.

The goal of this kind of study is primarily due to the hope of exploring the
possible existence of dynamical fluctuation (or intermittency) in high energy
collisions\r{2}. Remarkable success has been obtained recently in this respect
for hadron-hadron collisions --- experimental evidences for anisotropical
dynamical fluctuation (self-affine fractal) have been observed in $\pi^+$p
and $K^+$p collisions at 250 GeV/c\r{3}. It should be noticed,
however, that the ``intermittency'' phenomena in different kind of collision
processes may come from different physical origins, e.g.
QCD parton shower, Bose-Einstein correlation, second order phase
transition, etc. In some cases, especially in nuclear collision data
from emulsion experiments the influence of $\gamma$-conversion should
also be considered\r{4}. In the present paper we will not deal with the
physical origin of the phenomena and will concentrate on the comparison of
the phenomena in A-A and h-h collisions in order to study the characteristic
of nuclear effect and its possible application.

In comparing the scaling property of factorial moment in nucleus-nucleus and
hadron-hadron collisions, noticeable universal phenomena can be observed:

1) In one-dimension, the raise of the logarithm of FM (lnFM) as the
increasing of that of the phase space partition number $M$ is much weaker
for nucleus-nucleus than for hadron-hadron collisions, and the heavier 
the colliding nuclei are, the weaker is the rising of lnFM.

2) In higher-dimension, the lnFM for nucleus-nucleus collisions turns out to
be bending upwards strongly, much stronger than for hadron-hadron collisions, and the
heavier the colliding nuclei are, the stronger is the upward-bending of lnFM.

It has been shown in Ref.[5] that these two apparently contradictory facts
are both due to the superposition effect of the contribution from the large
number of elementary collisions in a nuclear collision process. This is a
geometrical effect and is valid regardless of what is the physical origin of
the phenomena. It may thus be useful in the search for new physics, such as
quark-gluon plasma (QGP)\r{5}.

In this paper these phenomena will be studied in more detail using the
nucleus-nucleus collision data from the EMU01 experiment.

\vskip 0.3cm
\noindent
{\bf II \ \ Method of analysis}
\vskip0.2cm

\noindent
Let us first recall briefly the superposition effect in the scaling behavior
of factorial moment in nuclear collisions\r{5}.

As is well known, in normal cases a nuclear collision process consists of
a large number of elementary nucleon-nucleon collisions.
Due to the complexity of the collision process,
the rapidity centers of individual elementary collisions
do not coincide but are scattered randomly within an interval on the rapidity
axis. For example, in central nuclear collision each participating nucleon in
an incident nucleus is facing a tube of nucleons in the other nucleus, so
that each nucleon will collide with a number of nucleons successively. The
difference in physical condition between these successive collisions makes
the rapidity centers of individual elementary collisions scattered on the
rapidity axis. When the (pseudo)rapidity region of each elementary collision
is divided into $M_{\parallel}$  pieces, their superposition makes the whole
rapidity region be divided into a much larger number ($M_{\parallel}^{\rm eff}$)
of pieces, cf. Fig.1.

On the other hand, similar effect does not exist in the transverse direction
($p_t$, $\varphi$) where the phase space region is the same for all the
elementary collisions, being (0,$p_t^{\rm max}$) for $p_t$ and (0,$2\pi$) for
$\varphi$, and their superposition makes no change in phase space partition.

A quantity $H$ called Hurst exponent can be used to characterize the way of
phase space partition. It is defined as
$$ H_{ab}={{\ln M_a}\over {\ln M_b}} \ , \eqno(2)$$
where $M_a$  and $M_b$ are the partition numbers in the directions $a$
and $b$ respectively. The anomalous scaling of FM, if exists, is definitely
connected with a certain value of H. If the way of phase space partition is
incorrect, that is to say, if the FM is not calculated with the right value of
$H$, the resulting ln$F_q$ vs. ln$M$ will be bending upwards\r{6}.

In nucleus-nucleus collisions, due to the superposition of elementary
collisions, the effective Hurst exponent
 $$ H_{\parallel\perp}^{\rm eff}={{{\rm ln} M_{\parallel}^{\rm eff}}\over
 {{\rm ln} M_{\perp}}} \gg {{{\rm ln}
   M_{\parallel}}\over {{\rm ln} M_{\perp}}}=H_{\parallel\perp} \ . \eqno(3) $$
Therefore, when we calculate FM with $H_{\parallel\perp}^{\rm cal}\approx
H_{\parallel\perp}\ll
 H_{\parallel\perp}^{\rm eff}$, we will observe that ln$F_q$ vs. ln$M$ is
bending strongly upwards. On the contrary, if we take a larger value for
$H_{\parallel\perp}^{\rm cal}$, the phenomenon of
upward bending should be weakened and eventually tend to vanish as
$H_{\parallel\perp}^{\rm cal}$ get close to $H_{\parallel\perp}^{\rm eff}$.

In order to characterize the degree of upward-bending we use a simple
quadratic function $y=ax^2$ to fit the ln$F_2$ vs. ln$M$ data\footnote{
In general, a linear term $bx$ should also be present in the function
used for fitting. However, in the present case, similar to the reasoning
leading to phenomenon 1 listed in Introduction, due to the large
number of nucleons in the colliding nuclei the coefficient $b$ of the
linear term, characterizing the strength of anomalous scaling, is very 
small and can therefore be omitted.}
In the fitting the first few data points have been omitted to reduce
the effect of momentum conservation\r{7}. The origin of coordinates is
then moved to the starting point ($M_0, F_2(M_0)$) of fitting and the
variables are changed correspondingly to
  $$ x=\ln M-\ln M_0 \ , \ y=\ln F_2(M)-\ln F_2(M_0). $$
The fitting value of $a$, being proportional to the second order
derivative $d^2y/dx^2$, is taken to be the characteristic parameter for the
degree of upward-bending of ln$F_2$ vs. ln$M$.

\vskip 0.3cm
\noindent
{\bf III \ \ The experimental setup}
\vskip0.2cm

\noindent
The EMU01 collaboration has collected data from collisions between various
projectiles and targets at different incident energies in the
ultra-relativistic
region\r{8}. Two different techniques have been employed, both utilizing nuclear
emulsion; ordinary emulsion stacks with exposures parallel to the emulsion plates
and emulsion chambers in which the exposures are perpendicular to the plates.
The second technique is best suited for an analysis in which the azimuthal
emission angles are of interest since, in this case, the detector has azimuthal
symmetry and high resolution. With this technique a resolution of
   $\Delta\eta\simeq 0.013$ rapidity
units in the central region can be obtained\r{9}. Pseudo-rapidity is given by
   $\eta =-\ln( \tan ( \theta /2))$, where $\theta$ is the emission angle with respect to the beam direction.
A major fraction of the
chambers are equiped with thin target foils, providing possibilities to study
interactions with various targets.
     In this paper, we have analysed data from interactions induced by the
CERN/SPS 200 A GeV Oxygen and Sulphur beams, 60 A GeV Oxygen beams and
the BNL/AGS 11A GeV Gold beams on targets of emulsion.
  The data samples used are as the following:\\

\begin{tabular}{|c|c|c|c|}\hline
 Experments  & Energy        &\multicolumn{2}{|c|}{Number of events}
  \\ \cline{3-4}
             & (A GeV)       & Mini-Bias Samples & Central Samples\\ \hline
 $^{197}$Au-Em      &    11         &                  & 45           \\ \hline
  $^{16}$O-Em       &    60         & 903              & 32             \\ \hline
  $^{16}$O-Em       &    200        & 730              &              \\ \hline
  $^{32}$S-Em       &    200        & 889              &              \\ \hline
\end{tabular}\\

Further details on the experiment, measurements and experimental criteria
can be found elsewhere\r{10}.

\vskip 0.3cm
\noindent
{\bf IV \ \ Results}
\vskip0.2cm

\noindent
We have analysed the second order FM for the experimental data
with several values of Hurst exponent
$(H=1.0, 0.8, 2.0, 3.0)$ in the plane $(\eta , \varphi)$.
Both central and minimum-bias events have been used.
The partition numbers in longitudinal and transverse directions
are chosen respectively as

\centerline{$ M_\varphi = 1,2,\dots,50, \ \ 
    M_\eta = M_\varphi^H$, for $H=1.0, \ 0.8$ ;}

\centerline{$ M_\eta = 1,2,\dots,50, \ \ 
    M_\varphi = M_\eta^{1/H}$, for $H=2.0, \ 3.0$ .}

\noindent The method proposed in Ref.[11] has been used for noninteger $M$.

In Fig's.2 and 3 the results for the minimum-bias data samples from
$^{32}$S-Em (200 A GeV) and $^{16}$O-Em (200  and 60 A GeV)  are shown respectively.

In Fig.4 the same analysis is repeated for central collision data samples from
11 A GeV $^{197}$\hskip-2truept Au-Em and 60 A GeV $^{16}$O-Em.

The pseudo-rapidity region used is
[$\eta_{\rm center}-2, \eta_{\rm center}+2$],
and the azimuthal region is [$0, 2\pi$]. In order to reduce the effect
of non-flat average distribution the cumulative variables $\chi _\eta$
and $\chi _\varphi$ have been used instead of $\eta$ and $\varphi$\r{12}.
The corresponding regions then become [0, 1].

It can be seen from the figures that when $H=1$, the curves bend upwards
strongly, and the smaller is the value of $H$, the more strongly is the
upward-bending. When $H$ increases, the upward-bending is weakened, and at
$H=2$ or $3$ the curves are almost
straight lines. This means that, in order to recover the anomalous scaling
of FM, the phase space should be divided finer in the longitudinal direction
than in the transverse direction. This is in contrary to the case of
hadron-hadron collisions, where the anomalous scaling of FM is obtained
with $H<1$\r{3}, i.e. when the phase space is divided finer in the transverse
direction than in the longitudinal direction. Since each elementary collision
process in the nuclear collision is expected to mimic hadron-hadron collision,
the finer partition in longitudinal direction for nuclear collisions could
only be due to the superposition effect shown schematically in Fig.1.

Next, we try to get a quantitative description of the phenomena by fitting
the upward-bending curves with the quadratic function $y=ax^2$ as described
in section II. In order to show the quality of fitting we plot in Fig.5
the results of fitting for the $H=1$ curve of $^{197}$\hskip-2truept Au-Em 
central collision (the first figure in Fig.4$a$).
In Fig.5$a$ all the points have been used and the origin is shifted to the
first point. In Fig.5$b,c,d$ the first 3,4,5 points are omitted and the origin
is shifted to the 4-, 5-, 6th point respectively. It can be
seen from the figures that the fit is good after omitting the first few
points and the result is insensitive to the exact number of omitted points
provided it is not too small, e.g. not smaller than 3.

The fit is unsatisfactory when all the points are included, cf. Fig.5$a$.
This is due to momentum conservation. The later tends to spread the particles
to opposite directions and thus reduce the value of FM. This effect becomes
weaker when $M$ increases. This explains why, when all the points are included,
the data points for medium values of $M$ lie below the fitting curve,
cf. Fig.5$a$.

In order to study the dependence of the above-mentioned superposition 
effect on the mass of colliding nuclei, we have compared the results from
central collision samples shown in Fig.4, where the impact parameter 
smearing is the minimum.

The fitting parameter $a$ for all the curves in Fig.4, with the first three
points omitted, are shown in Fig.6. From Fig.6 we see that $a$ decreases with
the increasing of $H$, i.e. the ln$F_2$ vs. ln$M$ curves bend upwards stronger
for smaller $H$, and tend to become straight lines as $H$ increases.

From Fig.6 we also find that the values of $a$ is always bigger 
for $^{197}$\hskip-2truept Au-Em
than for $^{16}$O-Em collisions at the same value of $H$.
This means that the ln$F_2$ vs. ln$M$ curves bend upwards stronger
for heavier colliding nuclei than for lighter ones at the same $H$.
This is just what to be expected, because for heavier colliding nuclei
the number of elementary collisions is larger, making the
$H_{\parallel\perp}^{\rm eff}$ bigger. When we calculate with the same
value of $H_{\parallel\perp}^{\rm cal}$ then this value will be further
away from $H_{\parallel\perp}^{\rm eff}$ for heavier
colliding nuclei than for lighter ones.

It can also be observed in Fig's.2--4 that for those values of $H>1$, which
make the ln$F_2$ vs. ln$M$ curves approximately straight, the slope of these
straight lines are small. This can also be understood from the superposition
effect in nucleus-nucleus collisions, which reduces the fluctuations
characterized by the slope of ln$F_2$ vs. ln$M$ (cf. the phenomenon 1 listed
in Introduction).

It should be stressed that the 2-D results from EMU01 have main
contribution from $\gamma$-conversion\r{4} and so can not be used to discuss
the physical property of the final state hadronic system. However, the
superposition effect discussed above is mainly a geometrical effect.
It concerns only the different ways of phase space division and does not
depend on the concrete property of the particles in consideration.
Therefore, the above discussion is appropriate for the 2-D data from
EMU01 as well.

\vskip 0.3cm
\noindent
{\bf IV \ \ Conclusion}
\vskip0.2cm

\noindent
We have investigated the 2-dimensional ($\eta, \varphi$) ln$F_2$ vs. ln$M$ with
different values of $H$ from both the minimum-bias and the central collision
data of EMU01 experiments. The value of $H_{\parallel\perp}^{\rm eff}$ in
nucleus-nucleus collision is found to be larger than unity. This means that
only when we use a right value of $H_{\parallel\perp}^{\rm cal}$, which is
larger than unity, in the analysis, can the superposition of the
contributions from elementary collisions in nucleus-nucleus collision be
correctly accounted for, and thus a straight line in ln$F_2$ vs. ln$M$
be obtained.

A parameter $a$ is introduced to characterize the degree of upward-bending
of ln$F_2$ vs. ln$M$. Using this parameter it is found that
the heavier are the colliding nuclei, the stronger is the upward-bending
of ln$F_2$ vs. ln$M$, in consistence with the fact that the number of
elementary collisions is larger for heavier colliding nuclei.

Our results show clearly the existence of superposition effect in the
behavior of higher-dimensional FM of nucleus-nucleus collisions and thus
support the argument proposed in Ref.[5] that the disappearence of
strongly upward-bending of higher-dimensional factorial moments may be
taken as a signal for the melting into a unique system of the produced
particles from elementary collisions in relativistic heavy ion collision.
\bigskip

\noindent
{\bf Acknowledgement}

This work is supported by NSFC, SECF and SCF of HuBei Province
China, NFR of Sweden, International Seminar in Uppsala, Minister of 
Research and Technology of Germany, TWRG, Department of University of 
Indian, DOE and NSF of USA. One of the authors (Liu L.S.) is grateful
to Dr. B. Wosiek and Wu Yuanfang for helpful discussions.

\newpage
\vskip2cm

\noindent{\bf \Large Figure Captions}
\bigskip
\begin{description}
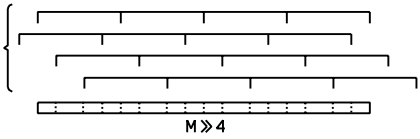
\item[Fig.1]
Schematic plot of the superposition effect on longitudinal phase space division.
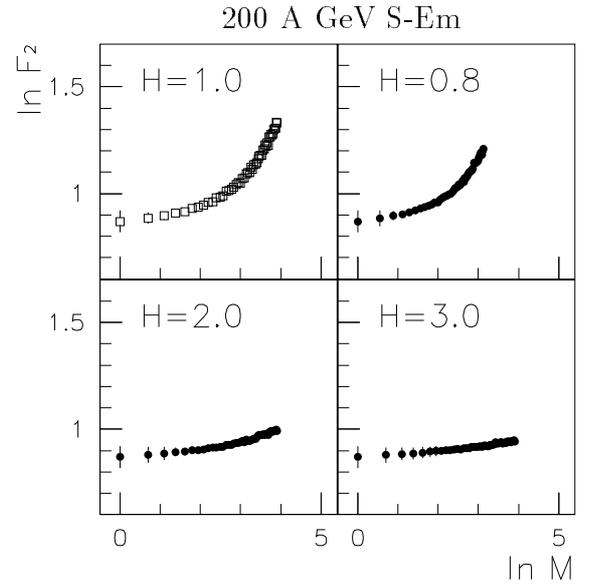
\item[Fig.2] Log-Log plot of 2-D ($\eta , \varphi$) $F_2$ vs. the partition
number $M(\eta)$ in longitudinal direction ($\eta$) from EMU01 200 A GeV
$^{32}$S-Em data for different values of $H$.
\item[Fig.3]
The same as Fig. 2 from 200 and 60 A GeV $^{16}$O-Em data.
\item[Fig.4]
The same as Fig. 2 from 11 A GeV $^{197}$\hskip-2truept Au-Em 
and 60 A GeV $^{16}$O-Em central collision data.
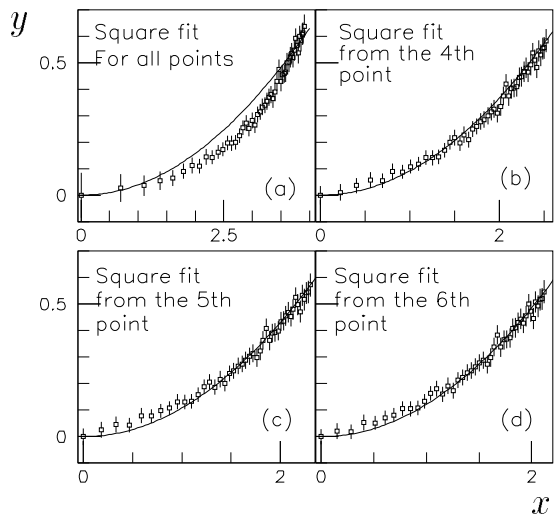
\item[Fig.5]
Fit the curve($H=1$) in Fig. 4$a$ by square function $y=ax^2$
after shifting the origin to the first, fourth, five, and six point.
Solid line stands for the fitting curve of $ax^2$
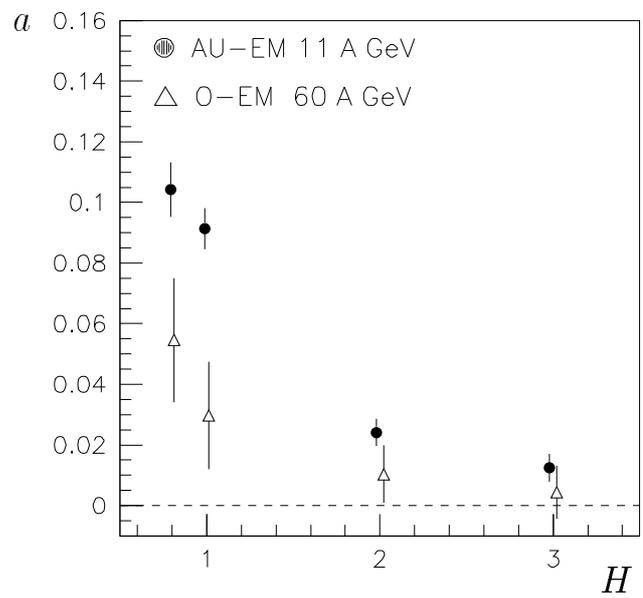
\item[Fig.6]
Comparison of the fitting parameter $a$ of the curves in Fig's.4 with
the origin of coordinate shifted to the fourth point.
\end{description}
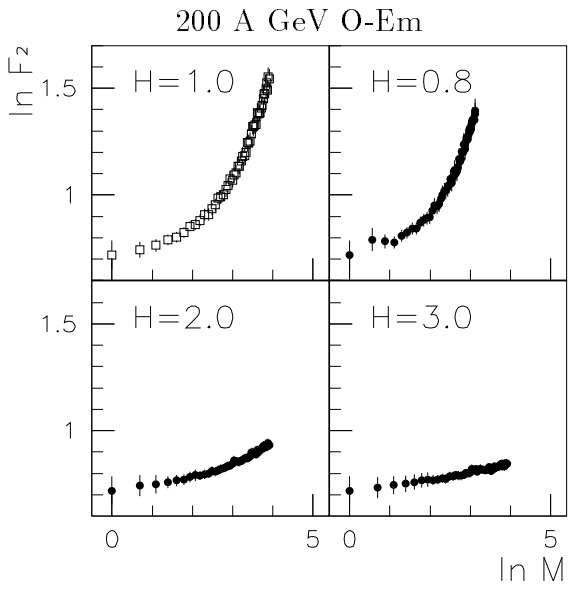
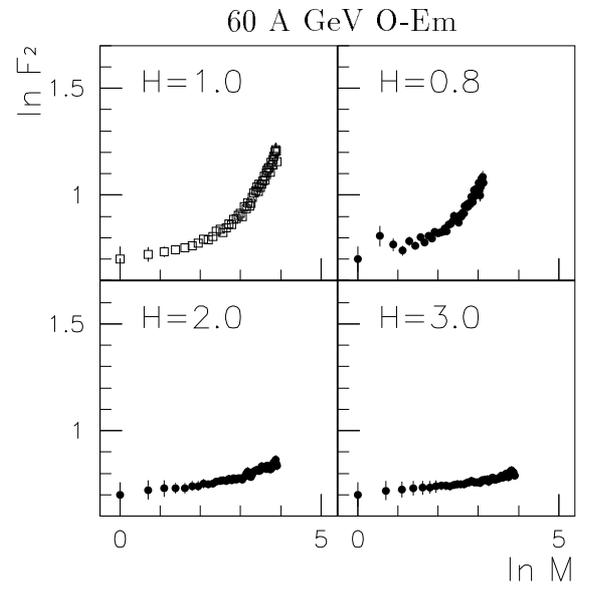
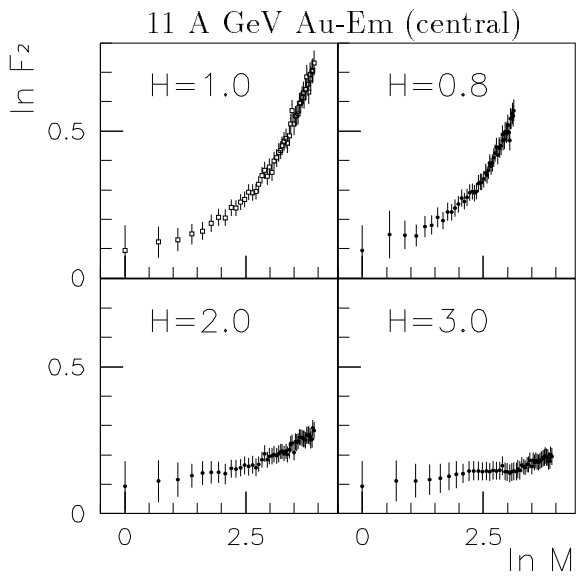
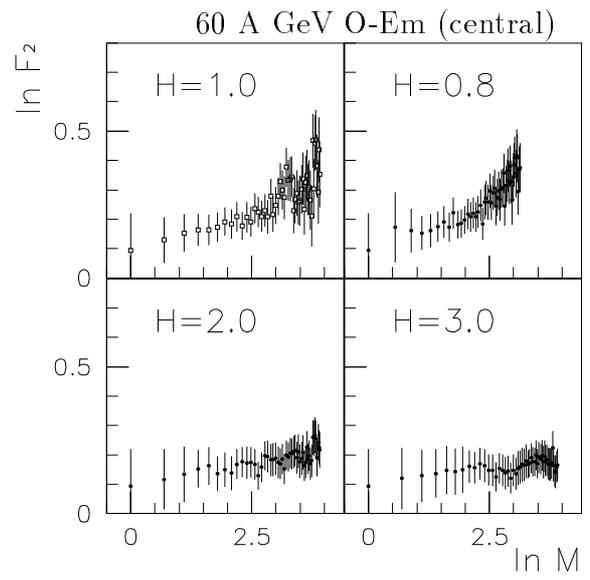
\end{document}